\documentclass[pra,twocolumn,floatfix,showpacs]{revtex4}
\usepackage{epsfig}
\usepackage{color}
\usepackage{amsmath}
\begin{document}                
\hyphenation{wave-guide}
\pagestyle{myheadings} \markright{3-Dimensional Mapping of Corneal
Topography and Thickness}
\title{3-Dimensional Mapping of Corneal Topography and Thickness}
\author{Jos\'e B. Almeida and Sandra Franco}
\affiliation{Universidade do Minho, Departamento de F\'isica,
4710-057 Braga, Portugal.} \email{bda@fisica.uminho.pt}



\date{\today}
\begin{abstract}                
An optical system that provides topographical maps of both
external and internal corneal surfaces as well as the latter's
thickness map is here described. Optical sections of the cornea
are obtained by illumination with a collimated beam expanded in a
fan shape by a small rotary cylindrical lens. The light diffused
from the cornea is observed by two cameras and processed in order
to yield the surfaces' profiles.

The optical system used to project a thin rotating line on the
cornea consists of a white light source provided with optical
fiber bundle output which is first conditioned by a set of lenses
so that it would produce a spot on the cornea. A small cylinder
lens is used to expand the beam in one direction so that a thin
line illuminates the cornea, rather than a spot. The cylinder lens
is provided with motor driven rotation about an axis normal to its
own in order to rotate the line on the cornea such that the
projected line scans the whole cornea; the illuminator is
completed with a slit aperture.

The cornea is not perfectly transparent, scattering some of the
light that traverses it; this fact is used for its observation by
two cameras. These are placed at pre-defined angles with the
illumination axis, so that optical sections of the cornea can be
seen; the use of two cameras avoids the need for camera rotation
in synchronism with the cylinder. The two cameras' images can be
combined in order to simulate a single virtual rotating camera.

Image processing is used to extract information about the corneal
surfaces profiles and thickness from the optical sections. Several
peculiar aspects of processing are discussed, namely the corneal
edge detection algorithm, the correction for angle of view and
deformation due to observation of the inner surface through the
cornea.

\end{abstract}

\maketitle
\section{Introduction}
With the advent of refractive surgery precise corneal pachymetry
thickness has become increasingly important, as parameters related
to corneal shape and thickness must be accurately measured in
order to ensure safety and accuracy ever more complex
interventions. Knowledge of corneal thickness is also useful when
studying corneal pathological conditions such as keratoconus,
\cite{Parafita00,Mandell69,Edmund87} investigating corneal
physiology \cite{Tomlinson72,Hage73} and in contact lens research.
\cite{Rivera96,Sanders75,Bonanno85,Holden85,Holden85:2}

Many pachymetry techniques have been developed for the
determination of central corneal thickness, with optical
pachymetry the most commonly used. However, this technique is slow
and the results are subjective because the operator must operate
the doubling device, estimating the point at which the lines
corresponding to the two corneal surfaces either just touch or
overlap, depending on the method used. On the other hand
ultrasonic pachymeters do not require much training and produce
more rapid and objective results, although they require an
anesthetized cornea, forbidden to some practitioners in some
countries. New pachymetric methods based on optical technology
have recently been developed and clinically applied, some of them
providing pachymetric maps and not only central thickness.
Confocal microscopy, videopachymetry, three-dimensional topography
marketed as Orbscan, optical tomography, non-contact and contact
specular microscopy, and low-coherence interferometry are other
techniques used in the measurement of corneal thickness. A review
of these methods has been published.\cite{Parafita02:3}

The authors developed an optical corneal tomographer that uses two
CCD cameras and an innovative illuminating system that allows
thickness measurements along any desired meridian and also
motorized scanning of the entire cornea. The optical principles
and technical details of a precursor apparatus have been
described. \cite{Franco00:2,Franco01,Franco02}
\section{System description}
\begin{figure}[tbh]
    \includegraphics[width=80mm]{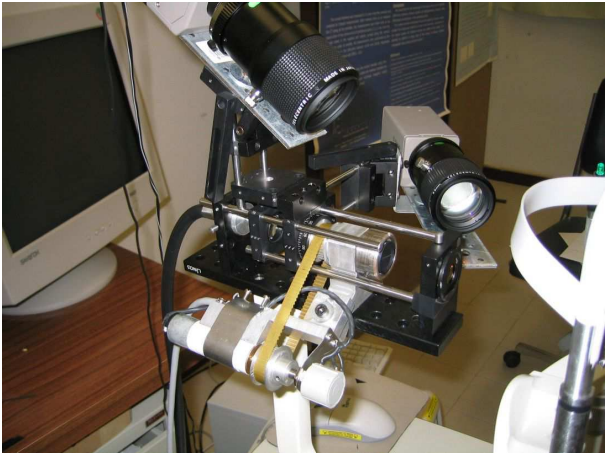}
    \caption{\label{fig:103_0328}View of the optical components;
    notice the two cameras and the illuminator.}
\end{figure}

The optical part of the system consists of an illuminator and two
CCD cameras (COU 2252) provided with $\text{55 mm}$ telecentric
lenses; this is complemented by an appropriate headrest and data
processing computer. Fig.\ \ref{fig:103_0328} shows a picture of
the optical components, where it is possible to see the special
arrangement of illuminator and the two cameras.

\begin{figure}[tbh]
    \includegraphics[width=80mm]{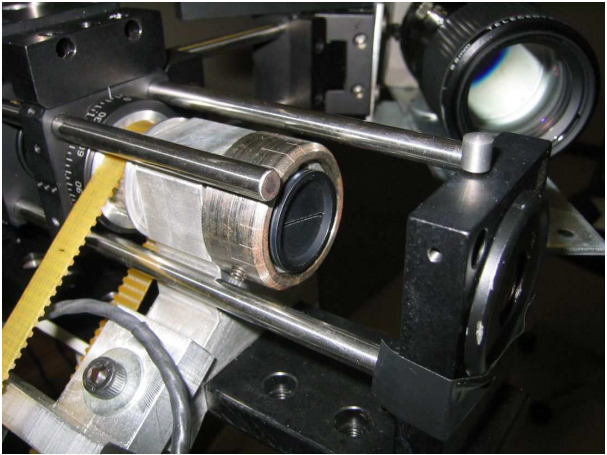}
    \caption{\label{fig:103_0329}Detailed view of the illuminator
    showing the rotation mechanism and the apodizing slit.}
\end{figure}

The illuminator is aligned with the visual axis and comprises a
quartz halogen light source an optical fiber bundle, a collimator,
a small cylindrical lens, a convex lens and an apodizing aperture
slit, see Fig.\ \ref{fig:103_0329}. After the beam is collimated,
a small cylindrical lens expands it in a fan. This lens has the
shape of a rod with a diameter of $\text{5 mm}$ and is held in a
mount that can be rotated to produce rotary scanning of all of the
cornea. The fan is then focused on the corneal surface by a convex
lens projecting a thin strip of light whose orientation follows
the cylinder lens' orientation.

\begin{figure}[tbh]
   \includegraphics[width=80mm]{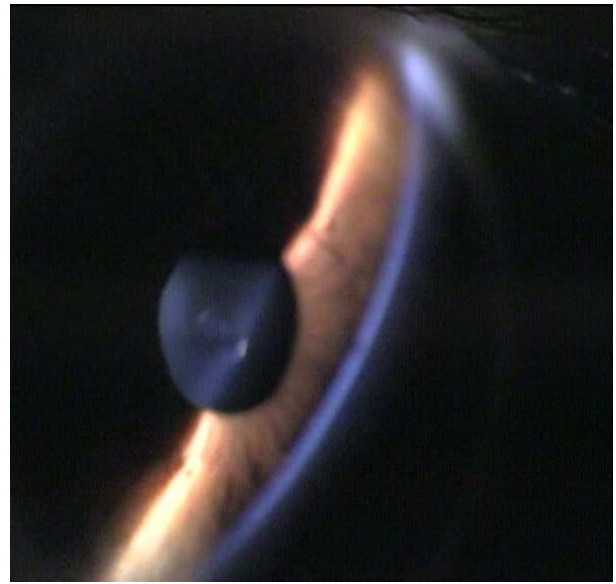}
    \caption{\label{fig:cornea3}Oblique section of the cornea obtained
    by the camera on the horizontal plane when the projected
    light strip makes an angle of approximately
     $45^\circ$ with the vertical
    direction.}
\end{figure}

The light diffused by both surfaces and the inner cornea is
collected by two video cameras placed at $60^\circ$ with the light
beam and defining with the visual axis planes perpendicular to
each other.
\begin{figure}[tbh]
   \includegraphics[width=80mm]{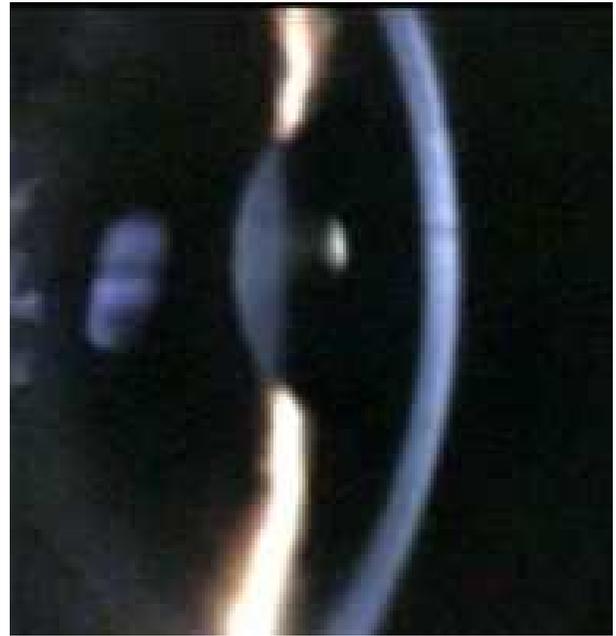}
    \caption{\label{fig:cornea1}Vertical section of an eye
     showing the cornea as a grey arc;
     the iris can be seen as the white saturated area with
     the lens sitting in the center.}
\end{figure}
Each camera sees an image similar to Fig.\ \ref{fig:cornea1},
where the cornea appears as a gray area in with the shape of an
arc; in the original colored image the cornea appears in dark blue
and can be easily distinguished from the iris, which is usually
shown in different color, frequently bright orange. Obviously the
arc's orientation depends on the camera and on the cylinder lens'
orientation, as shown in Fig.\ \ref{fig:cornea3}. The data of the
two cameras is then used to compute the corneal thickness. The two
cameras act like a single virtual camera that can be rotated with
the cylinder lens; the advantage of using two cameras is that it
allows faster rotation scanning then would be possible if a single
camera had to be rotated in synchronism with the lens.

\begin{figure}[tbh]
    \includegraphics[width=70mm]{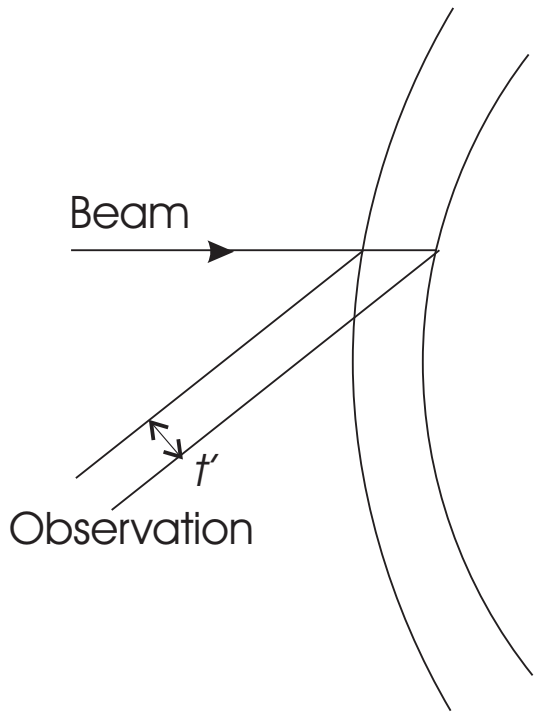}
    \caption{\label{fig:slitlamp}Working principle: the illuminating
    beam traverses the cornea and scattered light is observed at
    an established angle; the observed thickness $t'$ is apparent and
    must be corrected by software.}
\end{figure}

Figure \ref{fig:slitlamp} illustrates the working principle.
A light beam incident on the cornea is scattered on the inside and
the scattered light can be observed at an angle to the incidence
direction; since a corneal meridian is illuminated the scattered
light produces the arc shape already mentioned. The shape and
width of the arc have information about the corneal's shape and
thickness but data processing is needed to retrieve the correct
values for those parameters.
\section{Data processing}
A considerable amount of processing is necessary in order to
recover the true shape and thickness of the cornea from the
cameras' images we will detail the various steps below.
Fortunately processing speed is not important because all the
collected data can be saved for later processing, reducing
considerably the observation time.

The illustration in Fig.\ \ref{fig:cornea1} was obtained by the
camera laying on the horizontal plane when the projected light
strip was vertical; this image depicts a vertical section of the
cornea and the corresponding image on the vertical camera carries
no information because it is reduced to a bright straight line.
The situation is reversed when the projected strip is horizontal
but for all other situations there is information in both images,
which must be processed in order to obtain the image that would be
seen by a virtual camera on a plane always normal to the projected
strip. The first processing step consists in the application of
simple trigonometric rules to recover the virtual camera's image.

The second processing step consists in detecting the edges of the
gray arc image of the cornea using the method known as ''adaptive
thresholding'' reported by Hachicha et al. \cite{Hachicha89}. The
procedure involves the analysis of the gray-level along a scan
line crossing the corneal image and comparing this to a predefined
threshold level. A pixel $P_{i}$ is selected as an ''edge point,''
when it verifies the two simultaneous conditions:
\begin{equation}
\begin{split}
     \label{eq:threshold}
    I(P_{i}) &\geq I_{\min }+0.5(I_{\max }-I_{\min }) \\
    I(P_{i-1}) &< I_{\min }+0.5(I_{\max }-I_{\min })
\end{split}
\end{equation}
where the pixels are ranked from 1 to $N$ across the profile,
$I(P_{i}$) is the gray-level at pixel $P_{i}$ and $I_{min}$,
$I_{max}$ are the gray-level minimum and maximum, respectively.
Hachicha et al. \cite{Hachicha89} used a threshold of $0.3(I_{\max
}-I_{\min })$ in their work but we found that increasing the
factor from $0.3$ to $0.5$ gave us more reliable results. The
discrete edge points are then joined by spline fitting, allowing
determination of the normal direction and apparent thickness
measured along that direction.
\begin{figure}[tbh]
    \includegraphics[trim=110 0 110 0,width=80mm]{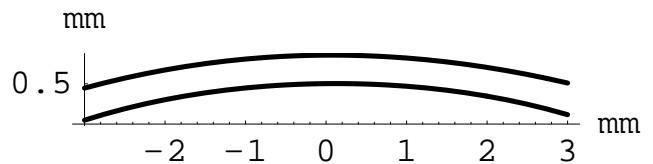}
    \caption{\label{fig:perfil}Corneal profile along a vertical meridian
     obtained by
    edge detection and spline fitting on the acquired image.}
\end{figure}

\begin{figure}[tbh]
    \includegraphics[trim=20 0 20 0,width=80mm]{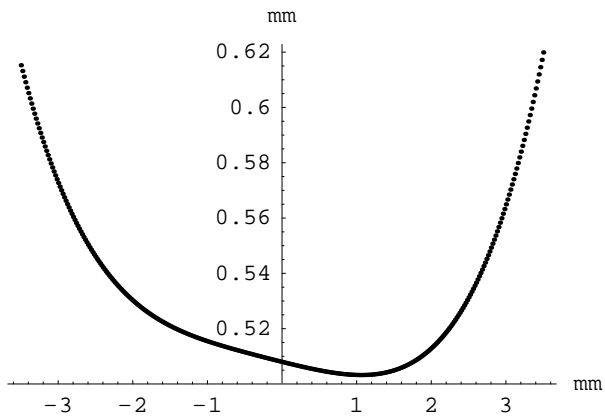}
    \caption{\label{fig:esp}Corneal thickness from the profile in
    Fig. \ref{fig:perfil} vs. distance from the
    visual axis.}
\end{figure}

The corneal thickness was computed from the distance between the
two edge profiles affected by corrections to allow for the
observation angle and corneal curvature. The former of these
corrections was detailed in previous work \cite{Franco02} and
consists only in dividing the apparent thickness by the sine of
the observation angle; the latter was performed considering the
optical magnification produced by an average curvature and could
be improved by iterative processing, using pre-determined
curvature at each point. We are presently developing the software
for data display in the form of topographical and pachymetric
maps; for the moment we can produce profile and thickness graphics
along any chosen meridian, as illustrated in Figs.\
\ref{fig:perfil} and \ref{fig:esp}.
\section{Conclusion}
The authors describe a novel optical system designed to provide
topographical maps of both corneal surfaces and corneal
pachymetry. Although not fully developed the system is already
capable of delivering clinically meaningful information, relevant
for diagnosis and surgery.

The authors describe the optical system used to project a thin
rotating line on the cornea. This is obtained with a white light
source provided with optical fiber bundle output; the light from
the fiber bundle is first conditioned by a set of lenses so that
it would produce a spot on the cornea. A small cylinder lens is
used to expand the beam in one direction so that a thin line
illuminates the cornea, rather than a spot. The cylinder lens is
provided with motor driven rotation about an axis normal to its
own in order to rotate the line on the cornea such that the
projected line scans the whole cornea.

The cornea is not perfectly transparent, scattering some of the
light that traverses it; this fact is used for its observation by
two cameras. These are placed at pre-defined angles with the
illumination axis, so that optical sections of the cornea can be
seen; the use of two cameras avoids the need for camera rotation
in synchronism with the cylinder. The two cameras' images can be
combined in order to simulate a single virtual rotating camera.

Image processing was used to extract information about the corneal
surfaces profiles and thickness from the optical sections. Several
peculiar aspects of processing were presented and discussed,
namely the corneal edge detection algorithm, the correction for
angle of view and deformation due to observation of the inner
surface through the cornea. Some examples of observed corneal
profiles were shown.

  \bibliography{Abrev,aberrations,assistentes}   

\begin{thebibliography}{16}
\expandafter\ifx\csname natexlab\endcsname\relax\def\natexlab#1{#1}\fi
\expandafter\ifx\csname bibnamefont\endcsname\relax
  \def\bibnamefont#1{#1}\fi
\expandafter\ifx\csname bibfnamefont\endcsname\relax
  \def\bibfnamefont#1{#1}\fi
\expandafter\ifx\csname citenamefont\endcsname\relax
  \def\citenamefont#1{#1}\fi
\expandafter\ifx\csname url\endcsname\relax
  \def\url#1{\texttt{#1}}\fi
\expandafter\ifx\csname urlprefix\endcsname\relax\def\urlprefix{URL }\fi
\providecommand{\bibinfo}[2]{#2}
\providecommand{\eprint}[2][]{\url{#2}}

\bibitem[{\citenamefont{Parafita et~al.}(2000)\citenamefont{Parafita,
  Gonz{\'a}lez-{M}{\'e}ijome, D{\'{\i}}az, and Yebra-{P}imentel}}]{Parafita00}
\bibinfo{author}{\bibfnamefont{M.~A.} \bibnamefont{Parafita}},
  \bibinfo{author}{\bibfnamefont{J.~M.}
  \bibnamefont{Gonz{\'a}lez-{M}{\'e}ijome}},
  \bibinfo{author}{\bibfnamefont{J.~A.} \bibnamefont{D{\'{\i}}az}},
  \bibnamefont{and}
  \bibinfo{author}{\bibfnamefont{E.}~\bibnamefont{Yebra-{P}imentel}},
  \bibinfo{journal}{Arch. Soc. Esp. Oftalmol.} \textbf{\bibinfo{volume}{75}},
  \bibinfo{pages}{633} (\bibinfo{year}{2000}).

\bibitem[{\citenamefont{Mandell and Polse}(1969)}]{Mandell69}
\bibinfo{author}{\bibfnamefont{R.~B.} \bibnamefont{Mandell}} \bibnamefont{and}
  \bibinfo{author}{\bibfnamefont{K.~A.} \bibnamefont{Polse}},
  \bibinfo{journal}{Arch. Ophthalmol.} \textbf{\bibinfo{volume}{82}},
  \bibinfo{pages}{182} (\bibinfo{year}{1969}).

\bibitem[{\citenamefont{Edmund}(1987)}]{Edmund87}
\bibinfo{author}{\bibfnamefont{C.}~\bibnamefont{Edmund}},
  \bibinfo{journal}{Acta Ophthalmol.} \textbf{\bibinfo{volume}{65}},
  \bibinfo{pages}{145} (\bibinfo{year}{1987}).

\bibitem[{\citenamefont{Tomlinson}(1972)}]{Tomlinson72}
\bibinfo{author}{\bibfnamefont{A.}~\bibnamefont{Tomlinson}},
  \bibinfo{journal}{Acta Ophthalmol.} \textbf{\bibinfo{volume}{50}},
  \bibinfo{pages}{73} (\bibinfo{year}{1972}).

\bibitem[{\citenamefont{El~{H}age and Beaulne}(1973)}]{Hage73}
\bibinfo{author}{\bibfnamefont{S.~G.} \bibnamefont{El~{H}age}}
  \bibnamefont{and} \bibinfo{author}{\bibfnamefont{C.}~\bibnamefont{Beaulne}},
  \bibinfo{journal}{Am. J. Optom. Physiol. Opt.} \textbf{\bibinfo{volume}{50}},
  \bibinfo{pages}{863} (\bibinfo{year}{1973}).

\bibitem[{\citenamefont{Rivera and Polse}(1996)}]{Rivera96}
\bibinfo{author}{\bibfnamefont{R.~K.} \bibnamefont{Rivera}} \bibnamefont{and}
  \bibinfo{author}{\bibfnamefont{K.~A.} \bibnamefont{Polse}},
  \bibinfo{journal}{Optom. Vis. Sci.} \textbf{\bibinfo{volume}{73}},
  \bibinfo{pages}{178} (\bibinfo{year}{1996}).

\bibitem[{\citenamefont{Sanders et~al.}(1975)\citenamefont{Sanders, Polse,
  Sarver, and Harris}}]{Sanders75}
\bibinfo{author}{\bibfnamefont{T.~L.} \bibnamefont{Sanders}},
  \bibinfo{author}{\bibfnamefont{K.~A.} \bibnamefont{Polse}},
  \bibinfo{author}{\bibfnamefont{M.~D.} \bibnamefont{Sarver}},
  \bibnamefont{and} \bibinfo{author}{\bibfnamefont{M.~G.}
  \bibnamefont{Harris}}, \bibinfo{journal}{Am. J. Optom. Physiol. Optics}
  \textbf{\bibinfo{volume}{52}}, \bibinfo{pages}{393} (\bibinfo{year}{1975}).

\bibitem[{\citenamefont{Bonanno and Polse}(1985)}]{Bonanno85}
\bibinfo{author}{\bibfnamefont{J.~A.} \bibnamefont{Bonanno}} \bibnamefont{and}
  \bibinfo{author}{\bibfnamefont{K.~A.} \bibnamefont{Polse}},
  \bibinfo{journal}{Am. J. Optom. Physiol. Opt.} \textbf{\bibinfo{volume}{62}},
  \bibinfo{pages}{74} (\bibinfo{year}{1985}).

\bibitem[{\citenamefont{Holden et~al.}(1985{\natexlab{a}})\citenamefont{Holden,
  Sweeney, Vannas, Nilsson, and Efron}}]{Holden85}
\bibinfo{author}{\bibfnamefont{B.~A.} \bibnamefont{Holden}},
  \bibinfo{author}{\bibfnamefont{D.~F.} \bibnamefont{Sweeney}},
  \bibinfo{author}{\bibfnamefont{A.}~\bibnamefont{Vannas}},
  \bibinfo{author}{\bibfnamefont{K.~T.} \bibnamefont{Nilsson}},
  \bibnamefont{and} \bibinfo{author}{\bibfnamefont{N.}~\bibnamefont{Efron}},
  \bibinfo{journal}{Invest. Ophthalmol. Vis. Sci.}
  \textbf{\bibinfo{volume}{26}}, \bibinfo{pages}{1489}
  (\bibinfo{year}{1985}{\natexlab{a}}).

\bibitem[{\citenamefont{Holden et~al.}(1985{\natexlab{b}})\citenamefont{Holden,
  Mc{N}ally, Mertz, and Swarbrick}}]{Holden85:2}
\bibinfo{author}{\bibfnamefont{B.~A.} \bibnamefont{Holden}},
  \bibinfo{author}{\bibfnamefont{J.~J.} \bibnamefont{Mc{N}ally}},
  \bibinfo{author}{\bibfnamefont{G.~W.} \bibnamefont{Mertz}}, \bibnamefont{and}
  \bibinfo{author}{\bibfnamefont{H.~A.} \bibnamefont{Swarbrick}},
  \bibinfo{journal}{Acta Ophthalmol.} \textbf{\bibinfo{volume}{63}},
  \bibinfo{pages}{684} (\bibinfo{year}{1985}{\natexlab{b}}).

\bibitem[{\citenamefont{Parafita et~al.}(2002)\citenamefont{Parafita,
  Yebra-Pimentel, Gir{\'a}ldez, Gonz{\'a}lez-P{\'e}rez,
  Gonz{\'a}lez-M{\'e}ijome, and Cervi{\~n}o}}]{Parafita02:3}
\bibinfo{author}{\bibfnamefont{M.~A.} \bibnamefont{Parafita}},
  \bibinfo{author}{\bibfnamefont{E.}~\bibnamefont{Yebra-Pimentel}},
  \bibinfo{author}{\bibfnamefont{M.~J.} \bibnamefont{Gir{\'a}ldez}},
  \bibinfo{author}{\bibfnamefont{J.}~\bibnamefont{Gonz{\'a}lez-P{\'e}rez}},
  \bibinfo{author}{\bibfnamefont{J.~M.}
  \bibnamefont{Gonz{\'a}lez-M{\'e}ijome}}, \bibnamefont{and}
  \bibinfo{author}{\bibfnamefont{A.}~\bibnamefont{Cervi{\~n}o}}, in
  \emph{\bibinfo{booktitle}{Recent Research Developments in Optics}}, edited by
  \bibinfo{editor}{\bibfnamefont{S.~G.} \bibnamefont{Pandalai}}
  (\bibinfo{publisher}{Research Signpost}, \bibinfo{address}{India},
  \bibinfo{year}{2002}), pp. \bibinfo{pages}{33--51}.

\bibitem[{\citenamefont{Franco et~al.}(2000{\natexlab{a}})\citenamefont{Franco,
  Almeida, and Parafita}}]{Franco00:2}
\bibinfo{author}{\bibfnamefont{S.}~\bibnamefont{Franco}},
  \bibinfo{author}{\bibfnamefont{J.~B.} \bibnamefont{Almeida}},
  \bibnamefont{and} \bibinfo{author}{\bibfnamefont{M.}~\bibnamefont{Parafita}},
  \bibinfo{journal}{J. Refract. Surg.} \textbf{\bibinfo{volume}{16}},
  \bibinfo{pages}{S661} (\bibinfo{year}{2000}{\natexlab{a}}),
  \bibinfo{note}{re-printed from \cite{Franco00:4}}.

\bibitem[{\citenamefont{Franco et~al.}(2001)\citenamefont{Franco, Almeida, and
  Parafita}}]{Franco01}
\bibinfo{author}{\bibfnamefont{S.}~\bibnamefont{Franco}},
  \bibinfo{author}{\bibfnamefont{J.~B.} \bibnamefont{Almeida}},
  \bibnamefont{and} \bibinfo{author}{\bibfnamefont{M.}~\bibnamefont{Parafita}},
  in \emph{\bibinfo{booktitle}{Vision Science and its Applications -- {VSIA}}}
  (\bibinfo{organization}{Optical Society of America},
  \bibinfo{address}{Monterey, California, USA}, \bibinfo{year}{2001}), pp.
  \bibinfo{pages}{148--151}.

\bibitem[{\citenamefont{Franco et~al.}(2002)\citenamefont{Franco, Almeida, and
  Parafita}}]{Franco02}
\bibinfo{author}{\bibfnamefont{S.}~\bibnamefont{Franco}},
  \bibinfo{author}{\bibfnamefont{J.~B.} \bibnamefont{Almeida}},
  \bibnamefont{and} \bibinfo{author}{\bibfnamefont{M.}~\bibnamefont{Parafita}},
  in \emph{\bibinfo{booktitle}{3rd International Congress of Wavefront Sensing
  and Aberration-Free Refractive Correction}} (\bibinfo{address}{Interlaken,
  Switzerland}, \bibinfo{year}{2002}), vol.~\bibinfo{volume}{18} of
  \emph{\bibinfo{series}{J. Refract. Surg.}}, pp. \bibinfo{pages}{S630--S633}.

\bibitem[{\citenamefont{Hachicha et~al.}(1989)\citenamefont{Hachicha, Simon,
  Samson, and Hanna}}]{Hachicha89}
\bibinfo{author}{\bibfnamefont{A.}~\bibnamefont{Hachicha}},
  \bibinfo{author}{\bibfnamefont{S.}~\bibnamefont{Simon}},
  \bibinfo{author}{\bibfnamefont{J.}~\bibnamefont{Samson}}, \bibnamefont{and}
  \bibinfo{author}{\bibfnamefont{K.}~\bibnamefont{Hanna}},
  \bibinfo{journal}{Comput. Vision Graphics Image Process.}
  \textbf{\bibinfo{volume}{47}}, \bibinfo{pages}{131} (\bibinfo{year}{1989}).

\bibitem[{\citenamefont{Franco et~al.}(2000{\natexlab{b}})\citenamefont{Franco,
  Almeida, and Parafita}}]{Franco00:4}
\bibinfo{author}{\bibfnamefont{S.}~\bibnamefont{Franco}},
  \bibinfo{author}{\bibfnamefont{J.~B.} \bibnamefont{Almeida}},
  \bibnamefont{and} \bibinfo{author}{\bibfnamefont{M.}~\bibnamefont{Parafita}},
  in \emph{\bibinfo{booktitle}{Vision Science and its Applications}}, edited by
  \bibinfo{editor}{\bibfnamefont{V.}~\bibnamefont{Lakshminarayanan}}
  (\bibinfo{publisher}{OSA}, \bibinfo{address}{Washington D.C.},
  \bibinfo{year}{2000}{\natexlab{b}}), vol.~\bibinfo{volume}{35} of
  \emph{\bibinfo{series}{Topics in Optics and Photonics Series}}, pp.
  \bibinfo{pages}{297--301}.

\end{thebibliography}

\end{document}